\documentclass[a4paper,11pt]{article}

\usepackage{jheppub}

\usepackage[T1]{fontenc}
\usepackage{dsfont}

\title{Gauge theories in anti-selfdual variables}

\author[a,b]{Marco Bochicchio}
\author[a,c]{Alessandro Pilloni}

\affiliation[a]{INFN sez. Roma 1\\Piazzale A. Moro 2, Roma, I-00185, Italy}
\affiliation[b]{Scuola Normale Superiore (SNS)\\Piazza dei Cavalieri 7, Pisa, I-56100, Italy}
\affiliation[c]{Dipartimento di Fisica `Sapienza' Universit\`a di Roma\\Piazzale A. Moro 2, Roma, I-00185, Italy}

\emailAdd{marco.bochicchio@roma1.infn.it}
\emailAdd{alessandro.pilloni@roma1.infn.it}

\abstract{Some years ago the Nicolai map, viewed as a change of variables from the gauge connection in a fixed gauge to the anti-selfdual part of the curvature, has been extended
by the first named author to pure Yang-Mills from its original definition in $\mathcal{N}=1$ supersymmetric Yang-Mills. We study here the perturbative one-particle irreducible effective action in the anti-selfdual variables of
any gauge theory, in particular pure Yang-Mills, \emph{QCD} and $\mathcal{N}=1$ supersymmetric Yang-Mills. We prove that the one-loop one-particle irreducible effective action of a gauge theory mapped to the anti-selfdual variables in any gauge is identical to the one of the original theory.
This is due to the conspiracy between the Jacobian of the change to the anti-selfdual variables and an extra functional determinant that arises from the non-linearity of the coupling of the anti-selfdual curvature to an external source in the Legendre transform that defines the one-particle irreducible effective action. Hence we establish the one-loop
perturbative equivalence of the mapped and original theories on the basis of the identity of the one-loop one-particle irreducible effective actions. Besides, we argue that the identity of the perturbative one-particle irreducible effective actions extends order by order in perturbation theory.}

\usepackage{hyperref}

\usepackage{slashed}
\DeclareMathOperator{\tr}{tr}

\DeclareMathOperator{\Det}{Det}

\DeclareMathOperator{\ad}{ad}

\newcommand{\nequalone}{$\mathcal{N}=1$~\emph{SUSY YM}}
\newcommand{\caldet}[1]{\mathcal{D}et_{#1}}
\newcommand{\overalpha}{\left.\mathds{1}_{\frac1{\alpha}}\right.}
\renewcommand{\epsilon}{\varepsilon}
\begin{document}

\maketitle
\flushbottom

\section{Introduction}
Many years ago Nicolai proved~\cite{Nicolai:1979nr,Nicolai:1980jc} that in any supersymmetric theory with unbroken supersymmetry there exists a change of variables that sets in Gaussian form the action that occurs in the partition function of the mapped theory.
After Nicolai proved this general result many examples were worked out in detail by Nicolai himself and by other authors. \par
In particular De Alfaro-Fubini-Furlan-Veneziano~\cite{deAlfaro:1984gw,deAlfaro:1984hb} discovered that the Nicolai map in $ \mathcal{N}=1$ supersymmetric Yang-Mills ({\nequalone}) is the change of variables from the gauge connection $A_{m}$ to the anti-selfdual (\emph{ASD}) part $F^-_{m n}$ of the gauge curvature $F_{mn}$ in the light-cone gauge.
The Jacobian of the map in the light-cone gauge cancels exactly the gluino determinant, thus making the \emph{ASD} curvature $F^-_{m n}$ in this gauge a free ultralocal Gaussian field. \par
This map is a well defined change of variables in function space since the $3$ components of the gauge fixed connection, i.e. the $4$ components of $A_m$ minus the gauge-fixing condition for the light-cone component $A_+=0$, are mapped
to the $3$ independent components of the \emph{ASD} field $F^-_{mn}$, i.e. the $6$ components of the curvature minus the $3$ \emph{ASD} conditions $F^-_{m n}=-\,{^*\!F}^-_{m n}$, with $ ^*$ the Hodge dual. \par
Nevertheless, for a long time it has been unclear whether the Nicolai map could reproduce perturbation theory. For example it was puzzling how a seemingly free partition function could reproduce the perturbative beta function of {\nequalone} or the exact Novikov-Shifman-Vainshtein-Zakharov (\emph{NSVZ}) beta function~\cite{Novikov:1983uc}.
In particular, nobody thought that the map could make sense without the cancellation of the determinants in the light-cone gauge, thus outside {\nequalone}~\cite{Floreanini:1985xs}. \par

However, in the last years new attention has been brought back to the Nicolai map by the first named author~\cite{Bochicchio:2003kw,Bochicchio:2008vt,Bochicchio:2009dh,Bochicchio:2011en,Bochicchio:2011np,Bochicchio:2012bj}.
Firstly, it was observed~\cite{Bochicchio:2011en} that in {\nequalone} the cancellation of the determinants in the partition function in the light-cone gauge occurs only up to zero modes, and therefore when zero modes arise there is an additional contribution to the beta function due to the Pauli-Villars regulator.
Moreover, it was discovered~\cite{Bochicchio:2011en} that the Nicolai map could be combined with localization techniques of cohomological nature to reproduce the well known result that the gluino condensate is exactly localized on configurations with vanishing \emph{ASD} curvature, i.e. instantons. In this way the exact \emph{NSVZ} beta function is correctly recovered~\cite{Bochicchio:2011en}.
With this non-perturbative result there is no argument against the consistency of the map.

Furthermore, the map to the \emph{ASD} variables has been extended from its original definition in {\nequalone} to pure~\emph{YM}~\cite{Bochicchio:2003kw,Bochicchio:2008vt,Bochicchio:2011en}, and as a consequence to any gauge theory that extends pure~\emph{YM}, as a change of variables from the gauge connection to the \emph{ASD} part of the curvature in any fixed gauge, in particular in the Feynman gauge~\cite{Bochicchio:2008vt,Bochicchio:2011en}.
The idea in~\cite{Bochicchio:2008vt,Bochicchio:2011en} is that the map is well-defined {\it per se} in every gauge-fixed theory, even if the cancellation of determinants does not occur, because of the matching of the local number of degrees of freedom. \par

Nevertheless, some skepticism remains about the Nicolai map in {\nequalone}, even more so in pure~\emph{YM}.

Until now the change to the \emph{ASD} variables has been employed to reproduce the correct beta function only in non-perturbative sectors
of {\nequalone}~\cite{Bochicchio:2011en} and of pure~\emph{YM}~\cite{Bochicchio:2008vt}. The problem of the perturbative equivalence is unsettled yet, since the non-perturbative techniques rely on the existence of zero modes around non-trivial
configurations, absent in perturbation theory, in order to reproduce the correct beta function. This problem is even more acute if certain large-$N$ arguments are applied to {\nequalone} mapped to the \emph{ASD} variables as in Shifman~\cite{Shifman:2011tr} which rely on planar diagrams for which zero modes play no role. Furthermore, even though changes of variables have been extensively studied in the simplest cases~\cite{Anselmi:2012aq}, there is no systematic study of a non-linear change of variables like the one
to the \emph{ASD} variables.

In this paper we intend to enforce the validity of the map to the \emph{ASD} variables by developing the perturbation theory in these variables.
Conceptually there is an infinite family of maps to the \emph{ASD} variables, parameterized by the gauge choice $\mathcal G(A)=0$ for any admissible gauge condition.
Yet we have discovered that computationally the most effective way to reproduce the standard one-particle irreducible (\emph{$1$PI}) effective action in the covariant $\alpha$-gauge is to perform the change of variables from the connection to the \emph{ASD} curvature completed by a fourth auxiliary scalar field $c$ defined by $ D_{n}(\tilde A) \delta A_{n} =c$. $c$ is distributed as a Gaussian ultralocal field with covariance proportional to $\alpha$ and $D_{n}$ is the covariant derivative depending on the background field $\tilde A_n$ that occurs in the
construction of the usual \emph{$1$PI} effective action by the background field method. \par
Therefore, in the covariant $\alpha$-gauge we map the $4$ components of the gauge connection $A_m$ to the $3$ \emph{ASD} fields plus the field $c$. Thus the number of local degrees of freedom in the gauge-fixed functional integral is exactly conserved by our map. The map is local in the direction from the connection to the \emph{ASD} field but its inverse is non-local.
Moreover, the spin content of the original and mapped theories is the same.
Indeed, the 4-vector $A_m$ transforms as $\left(1/2,1/2\right)$ for $SU(2) \times SU(2)$ in Euclidean space-time and therefore decomposes into a 3-vector and a scalar, which can be eliminated by the gauge fixing in the Landau gauge $\alpha=0$.
The \emph{ASD} tensor transforms as $\left(0,1\right)$ and therefore its independent components coincide with the 3-vector $\vec{E}-\vec{H}$.

The plan of the paper is as follows. \par
In Section~\ref{sec:Conclusions} we add some further comments and we state our conclusions. \par
In Section~\ref{sec:asdvariables} we define the map to the \emph{ASD} variables in the \emph{YM} theory in any gauge, in particular in the covariant $\alpha$-gauge, and we evaluate the generating functional of perturbation theory employing the background field method. We remark the changes that occur in the \emph{$1$PI} effective action in the \emph{ASD} variables as opposed to the original variables, and we show that the coupling to an external source provides a contribution to the beta function additional to the Jacobian of the change of variables.
This is due to the non-linearity of the coupling of the source to the \emph{ASD} field induced via the change of variables by the linear coupling of the source to the gauge connection. \par
In Section~\ref{sec:effective} we prove that the one-loop \emph{$1$PI} effective action in the \emph{ASD} variables in any covariant $\alpha$-gauge is identical to the original one. In order to do so we employ a chiral
spinor notation. \par
In Section~\ref{sec:geometric} we extend the proof of the identity of the effective actions to any gauge $\mathcal G(A)=0$ (e.g. the light-cone gauge). This is obtained employing a differential geometric notation in terms of differential forms and of the Hodge dual. \par
In Section~\ref{sec:feynman} we check by direct computation that the one-loop beta function of the \emph{YM} theory mapped to the \emph{ASD} variables is identical to the one of the original theory
in the Feynman gauge. In this gauge the spinor notation makes the computation easier.
In intermediate steps we compute the Jacobian of the map to the \emph{ASD} variables in the Feynman gauge and the associated divergent counterterm, finding perfect agreement with the previous computation~\cite{Bochicchio:2008vt,Bochicchio:2011en}. \par

\section{Conclusions}
\label{sec:Conclusions}
In this paper we prove that the change to the \emph{ASD} variables in any gauge-fixed theory that extends pure~\emph{YM} is well defined, and it maps identically the one-loop \emph{$1$PI} effective actions
of the original and mapped theories. We work out in detail the case of pure~\emph{YM}, but the extension to $QCD$ or to {\nequalone} is immediate, as well as to any gauge theory involving the \emph{YM} connection.
Indeed, both in the original variables and in the $ASD$ variables, to obtain the one-loop effective action of these more general theories, it is sufficient to add to the $YM$ effective action
minus/plus the logarithm of the corresponding fermion/boson functional determinant in the background of the gauge field. \par
One may wonder as to whether in the \emph{YM} theory mapped to the \emph{ASD} variables chiral gauge anomalies may arise. Were to arise gauge anomalies, the Jacobian would be ill-defined, and our construction would be meaningless. However, the chiral determinant that is the Jacobian of the map to the \emph{ASD} variables is a matrix with indices in the adjoint representation of Lie algebra, that cannot give rise to anomalies being a real representation. \par
This is also why in {\nequalone} gluinos occur without introducing any anomaly, despite their contribution to the effective action is a chiral determinant, as it is the Jacobian to the \emph{ASD} variables. In fact, in the Feynman gauge the Jacobian is formally the square of the gluino determinant, because the gluon field, being a vector, has one more spinor index with respect to the gluino field. \par
As a consequence of the one-loop result of this paper there is no obstruction to extend the change to the \emph{ASD} variables order by order in perturbation theory. Indeed, if the change of variables is well defined,
i.e. invertible, at a certain order of perturbation theory, it is automatically invertible at the next order provided the gauge coupling is small enough. \par
The formulation of gauge theories in the \emph{ASD} variables, that is the main conceptual byproduct of this paper, shows that the gauge invariant particle content of \emph{YM} theories may in fact be very different from the one
inferred by conventional perturbation theory in terms of the gauge connection. Indeed, neither the gauge connection nor the \emph{ASD} variables are gauge invariant, but because of the equivalence of the two formulations they support the same gauge invariant particle content. \par
However, the \emph{ASD} variables are gauge covariant. Hence their eigenvalues are gauge invariant, a feature that has importance in non-perturbative applications.
In fact, the change to the \emph{ASD} variables has been the basis of new non-perturbative approaches to \emph{YM} and \emph{SUSY~YM}~\cite{Bochicchio:2008vt,Bochicchio:2009dh,Bochicchio:2011en,Bochicchio:2011np,Bochicchio:2012bj}.
More generally, it is a new language to describe gauge theories whose potentiality remains to be fully explored.

\section{Change to the \emph{ASD} variables}
\label{sec:asdvariables}
The generating functional of \emph{YM} in Euclidean space-time, omitting momentarily the gauge fixing, is:
\begin{equation}
Z\left[J\right]=\int \mathcal{D}A\, \exp \left[-\frac{1}{2g^2} \int d^4 x \, \tr \left(F_{mn}\right)^2 + \int d^4 x \,2\tr\,J_m A_m\right]
\end{equation}
where $A_m$ is the gauge connection and $F_{mn} = \partial_{[m,} A_{n]} + i \left[A_m,A_n\right]$ is the gauge curvature. $F_{mn},A_m,J_m$ are matrices valued in the fundamental representation of the Lie algebra of $SU(N)$, e.g. $A_m = A_m^a T^a$, with $\tr(T^a T^b) = \tfrac12 \delta^{ab}$. The following well known identity holds:
\begin{equation} \label{11}
\tr F_{mn}^2 = \frac12 \tr \left(F^{-}_{mn}\right)^2 + \tr F_{mn} \,{^*\!F}_{mn}
\end{equation}
with ${^*\!F}_{mn} =\frac{1}{2} \epsilon_{mnrs} F_{rs}$, $F^\pm = 2 P^\pm F=F \pm \,{^*\!F}$ and $P^\pm$ the projectors over the \emph{SD} (resp. \emph{ASD}) curvatures defined in Eq.~\eqref{5}. Thus:
\begin{equation}
\begin{split}
Z\left[J\right]&=\int \mathcal{D}A\, \exp \left[-\frac{1}{2g^2} \int d^4 x \, \tr F_{mn}\,{^*\!F}_{mn} -\frac{1}{4g^2} \int d^4 x \, \tr \left(F_{mn}^-\right)^2 + \int d^4 x \,2\tr\,J_m A_m\right]\\
&=\int \mathcal{D}A\, \exp \left[-\frac{\left(4\pi\right)^2 Q\left(A\right)}{2g^2} -\frac{1}{4g^2} \int d^4 x \, \tr \left(F_{mn}^-\right)^2 + \int d^4 x \,2\tr\,J_m A_m\right]
\end{split}
\end{equation}
where $Q=\tr \int d^4x F\, {^*\!F} / \left(4\pi\right)^2$ is the second Chern class. We set $Q=0$ since only trivial bundles occur in perturbation theory. \par
We now change variables from the connection to the \emph{ASD} curvature following~\cite{Bochicchio:2008vt}, inserting in the functional integral the appropriate resolution of the identity introduced in~\cite{Bochicchio:2003kw}:
\begin{equation}
1=\int \mathcal{D}\mu^-\, \delta\!\left(F_{mn}^--\mu_{mn}^-\right)
\end{equation}
The resolution of identity is an essential technical tool to perform explicit computations in any gauge.
Integrating the delta function on the gauge connection, we get formally the Jacobian, that is ill-defined because we have not fixed the gauge yet:
\begin{equation}
\begin{split}
Z\left[J\right]&=\int \mathcal{D}A\, \mathcal{D}\mu^-\,\exp \left[-\frac{1}{4g^2} \int d^4 x \, \tr \left(F_{mn}^-\right)^2 + \int d^4 x \,2\tr\,J_m A_m\right]\,\delta\!\left(F_{mn}^--\mu_{mn}^-\right)\\
&=\int \mathcal{D}\mu^-\,\exp \left[-\frac{1}{4g^2} \int d^4 x \, \tr \left(\mu_{mn}^-\right)^2 + \int d^4 x \,2\tr\,J_m A_m\right]\,\Det\frac{\delta A}{\delta \mu^-}
\end{split}
\end{equation}
Inserting the gauge-fixing condition in the covariant $\alpha$-gauge, the generating functional reads:
\begin{equation}
\begin{split}
Z\left[J\right]&=\int \mathcal{D}A\, \mathcal{D}\mu^- \mathcal{D}c \,\delta\!\left(F_{mn}^--\mu_{mn}^-\right) \, \delta \left(D_m(\tilde A) A_m -c\right) \, \Delta_\textrm{FP}\\
& \quad\exp \left[-\frac{1}{4g^2} \int d^4 x \, \tr \left(F_{mn}^-\right)^2 - \frac{1}{\alpha g^2} \int d^4 x \, \tr(c^2) + \int d^4 x \,2\tr\,J_m A_m\right]
\end{split}
\end{equation}
where $\Delta_\textrm{FP}$ is the Faddeev-Popov (\emph{FP}) determinant and the covariant derivative acts by the commutator \mbox{$D_m(\tilde A)=\partial_m +i [\tilde A_m, \cdot]$} and it is computed with respect to a background field $\tilde A_m$. \par
We find convenient at this stage to introduce a spinor notation in Euclidean space-time. To avoid confusion, we use latin characters for vector indices and greek characters for spinor indices:
\begin{subequations}
\label{eq:spinordefinitions}
\begin{align}
A_{\alpha\dot\alpha} &=A_m \left(\sigma^m\right)_{\alpha\dot\alpha}& \bar A^{\dot\alpha\alpha} &=A_m \left(\bar\sigma^m\right)^{\dot\alpha\alpha}\\
D_{\alpha\dot\alpha} &=D_m \left(\sigma^m\right)_{\alpha\dot\alpha}& \bar D^{\dot\alpha\alpha} &=D_m \left(\bar\sigma^m\right)^{\dot\alpha\alpha}\\
{\left(\mu^-\right)^{\dot\alpha}}_{\dot\beta} &= \frac12 \left(\mu^-\right)_{mn} {\left(\bar\sigma^{mn}\right)^{\dot\alpha}}_{\dot\beta} & {\left(\nu^-\right)^{\dot\alpha}}_{\dot\beta} &= {\left(\mu^-\right)^{\dot\alpha}}_{\dot\beta} + c\, \delta^{\dot\alpha}_{\dot\beta}\label{eq:mumenospinor}
\intertext{with:}
\left(\sigma^m\right)_{\alpha\dot\alpha} &=\left(\mathds{1},i \vec\tau\right)_{\alpha\dot\alpha} & \left(\bar\sigma^m\right)^{\dot\alpha\alpha} &=\left(\mathds{1},-i \vec\tau\right)^{\dot\alpha\alpha}
\end{align}%
and:
\begin{align}
{\left(\sigma^{mn}\right)_{\alpha}}^{\beta} &=\frac14\left[\left(\sigma^m\right)_{\alpha\dot\alpha}\left(\bar \sigma^n\right)^{\dot\alpha\beta} - \left(\sigma^n\right)_{\alpha\dot\alpha}\left(\bar\sigma^m\right)^{\dot\alpha\beta}\right]\\
{\left(\bar\sigma^{mn}\right)^{\dot\alpha}}_{\dot\beta} &=\frac14\left[\left(\bar\sigma^m\right)^{\dot\alpha\alpha}\left(\sigma^n\right)_{\alpha\dot\beta} - \left(\bar\sigma^n\right)^{\dot\alpha\alpha}\left(\sigma^m\right)_{\alpha\dot\beta}\right]
\end{align}
where $\vec\tau$ are the Pauli matrices and $\nu^-$ is a new field that includes the \emph{ASD} field $\mu^-$ and the longitudinal auxiliary field $c$. It follows that:
\begin{align}
\left(\sigma^m\right)_{\alpha\dot\alpha}\left(\bar \sigma^n\right)^{\dot\alpha\beta} + \left(\sigma^n\right)_{\alpha\dot\alpha}\left(\bar\sigma^m\right)^{\dot\alpha\beta}&= 2\delta^{mn} \delta_{\alpha}^{\beta}\\
\left(\bar\sigma^m\right)^{\dot\alpha\alpha}\left(\sigma^n\right)_{\alpha\dot\beta} + \left(\bar\sigma^n\right)^{\dot\alpha\alpha}\left(\sigma^m\right)_{\alpha\dot\beta} &= 2\delta^{mn} \delta^{\dot\alpha}_{\dot\beta}
\end{align}
\end{subequations}
Using Eqns.~\eqref{eq:spinordefinitions}, we write the map to the \emph{ASD} variables in spinor notation (spinor indices are understood):
\begin{align}
\mu^{-a}_{mn} &= 2 P^-_{mn,rs} \left(\partial_{[r,} A^a_{s]}-f^{abc} A^b_r A^c_s\right)\nonumber\\
\frac12 \bar \sigma^{mn} \mu^{-a}_{mn} &= \bar \sigma^{mn} \left(\partial_{[m,} A^a_{n]}-f^{abc} A^b_m A^c_n\right)\nonumber\\
\mu^{-a} &= \frac12 \left(\bar \sigma^{m} \sigma^n - \bar\sigma^n\sigma^m\right) \left(\partial_{m} A^a_{n}- \frac12 f^{abc} A^b_m A^c_n\right)\nonumber\\
\mu^{-a} &= \left(\bar \sigma^{m} \sigma^n - \frac12\bar\sigma^{\{m,}\sigma^{n\}}\right) \left(\partial_{m} A^a_{n}- \frac12 f^{abc} A^b_m A^c_n\right)\nonumber\\
\mu^{-a} &= \left(\bar \sigma^{m} \sigma^n - \delta_{mn} \mathds{1} \right) \left(\partial_{m} A^a_{n}- \frac12 f^{abc} A^b_m A^c_n\right)\nonumber\\
\mu^{-a} &= \bar \partial A^a - \mathds{1} \partial_m A^a_m - \frac12 f^{abc} \bar A^b A^c\label{eq:mapvero}
\end{align}
Moreover, in the covariant $\alpha$-gauge the map to the \emph{ASD} field $\mu^-$ and the gauge-fixing condition $c=D_m(\tilde A) A_m$ combine into:
\begin{equation}
\nu^{-a} = \bar \partial A^a - \frac12 f^{abc} \bar A^b A^c - \mathds{1} f^{abc} \tilde A^b_m \delta A^c_m
\end{equation}
Evaluating the fluctuation around the classical configuration $\tilde A$, we get:
\begin{equation}
\begin{split}
\delta\nu^{-a} &= \bar \partial \delta A^a - \frac12 f^{abc} \overline{ \delta A}^b \tilde A^c - \frac12 f^{abc} \overline{\tilde A}^b \delta A^c - \mathds{1} f^{abc} \tilde A^b_m \delta A^c_m   - \frac12 f^{abc} \overline{\delta A}^b \delta A^c \\
&=\bar \partial \delta A^a - \frac12 f^{abc} \left(-\overline{ \tilde A}^c \delta A^b + 2 \,\mathds{1} \delta A^b_m \tilde A^c_m\right) - \frac12 f^{abc} \overline{\tilde A}^b \delta A^c - \mathds{1} f^{abc} \tilde A^b_m \delta A^c_m   - \frac12 f^{abc} \overline{\delta A}^b \delta A^c \\
&= \bar \partial \delta A^a - f^{abc} \overline{ \tilde A}^b \delta A^c - \mathds{1} f^{abc} \delta A^b_m \tilde A^c_m - \mathds{1} f^{abc} \tilde A^b_m \delta A^c_m  - \frac12 f^{abc} \overline{\delta A}^b \delta A^c  \\
&= \bar D(\tilde A)^{ac} \delta A^c - \frac12 f^{abc} \overline{\delta A}^b \delta A^c = \bar D(\tilde A)^{ac} \delta A^c + O\left(\delta A^2\right)
\end{split}
\label{eq:mapnu}
\end{equation}
The Jacobian of the map depends only on the linear part of Eq.~\eqref{eq:mapnu}, hence it is now well defined since the operator $\bar D$ is invertible in perturbation theory. Therefore, the Jacobian is computed by:
\begin{equation}
\Det \frac{\delta A}{\delta\nu^-} =\int \mathcal{D}A \; \delta\!\left(\nu^{-a} - \bar \partial A^a + \frac12 f^{abc} \bar A^b A^c + \mathds{1} f^{abc} \tilde A^b_m \delta A^c_m\right)= \Det( \bar D)^{-1}
\end{equation}
Thus in the covariant $\alpha$-gauge in spinor notation the generating functional reads:
\begin{multline}
\label{ASD}
Z\left[J\right]=\int \mathcal{D}\nu^-\,\exp \Bigg[-\frac{1}{4g^2} \int d^4 x \, {\left(\nu^{-a}\right)^{\dot\alpha}}_{\dot\beta}\left[\delta^{\dot\gamma}_{\dot\alpha}\delta^{\dot\beta}_{\dot\delta} - \frac12\left(1-\frac1{\alpha}\right)\delta^{\dot\beta}_{\dot\alpha}\delta^{\dot\delta}_{\dot\gamma}\right]{\left(\nu^{-a}\right)_{\dot\gamma}}^{\dot\delta}\\
+ \int d^4 x\, \left(\bar J^a\right)^{\dot \alpha \alpha} \,\left[A\!\left(\nu^-\right)\right]^a_{\alpha\dot\alpha} \Bigg]\Det \frac{\delta A}{\delta\nu^-}\Det(-\Delta)
\end{multline}
with ${\left(\mu^{-a}\right)_{\dot\alpha}}^{\dot\beta} = \epsilon_{\dot\alpha\dot\gamma} \,\epsilon^{\dot\delta\dot\beta}\,{\left(\mu^{-a}\right)^{\dot\gamma}}_{\dot\delta}$
and $ \Det(-\Delta )$ the \emph{FP} determinant in the covariant $\alpha$-gauge, where $ \Delta = D^2$ is the Laplacian constructed by means of the covariant derivative. Writing the color indices explicitly:
\begin{equation}
( \Delta)^{ac}=({D}_m )^{ad}({D}_m)^{dc}= \partial^2 \delta^{ac}-\partial_m \tilde A^b_m f^{abc}- 2 \tilde A^b_m f^{abc}\partial_m + \tilde A^b_m f^{abd} \tilde A^e_m f^{dec}
\end{equation}
 \par
In Section~\ref{sec:effective} we evaluate the functional integral in  Eq.~\eqref{ASD} at one-loop.
Because of the nonlinear coupling of the source $J$ to $\nu^-$ induced by the linear coupling to the connection $A$, the source term at one loop provides a functional determinant in addition to the Jacobian of the change of variables. \par

In general gauges $\mathcal{G}\left[A\right]=0$ we simply restrict the map~\eqref{eq:mapvero} to the gauge-fixed slice of the gauge orbits. The corresponding generating functional reads:

\begin{multline}
Z\left[J\right]=\int \mathcal{D}\mu^-\,\exp \Bigg[-\frac{1}{4g^2} \int d^4 x \, {\left(\mu^{-a}\right)^{\dot\alpha}}_{\dot\beta}{\left(\mu^{-a}\right)_{\dot\alpha}}^{\dot\beta}
+ \int d^4 x\, \left(\bar J^a\right)^{\dot \alpha \alpha} \,\left[A\!\left(\mu^-\right)\right]^a_{\alpha\dot\alpha} \Bigg]\\ \Det \frac{\delta A}{\delta \mu^-}\,\Delta_\textrm{FP}\, \delta\left(\mathcal{G}\left[A\left(\mu^-\right)\right]\right)
\end{multline}
As in the covariant $\alpha$-gauge, because of the nonlinear coupling of the source $J$ to $\mu^-$, the source term at one loop provides a functional determinant in addition to the Jacobian of the change of variables. We evaluate it in Section~\ref{sec:geometric}. \par
We observe that
in the usual variables, as opposed to the \emph{ASD} variables, the source term does not provide any contribution to the \emph{$1$PI} effective action. This is easily seen evaluating at one loop the effective action $\Gamma$ as the Legendre transform of the connected generating functional $G$:
\begin{equation} \label{bf}
\exp\left(-\Gamma[\tilde{A}]\right) = \exp\left(-G\left[J\right]-J\tilde{A}\right)= Z\left[J\right]\exp(-J\tilde{A})= \int \mathcal{D}A \, e^{- \frac{1}{2g^2} S_{\textrm{YM}}\left[A\right] + J A - J \tilde{A}}
\end{equation}
where the sum on the discrete indices and the integrals on space-time are understood. $J=\frac{1}{2g^2} \left.\frac{\delta S_{\textrm{YM}}}{\delta A}\right|_{A=\tilde{A}}$, i.e. $\tilde A$ satisfies the equation of motion with the external source~$J$. Indeed, expanding $A = \tilde{A}+\delta A$ around the saddle point $\tilde{A}$:
\begin{equation}
\begin{split}
\exp\left(-\Gamma[\tilde{A}]\right) &= \int \mathcal{D} \delta A \, \exp\left(- \frac{1}{2g^2}S_{\textrm{YM}}[\tilde{A}+\delta A] + J \,\delta A\right)\\
&\sim e^{-\frac{1}{2g^2}S_{\textrm{YM}}\left[\tilde{A}\right]} \int \mathcal{D} \delta A \, \exp\left(- \frac{1}{2g^2} \left.\frac{\delta S_{\textrm{YM}}}{\delta A}\right|_{\tilde{A}} \delta A + J \,\delta A - \frac{1}{4g^2}\left.\frac{\delta^2 S_{\textrm{YM}}}{\delta A^2}\right|_{\tilde A} \delta A^2\right)\\
&= e^{-\frac{1}{2g^2}S_{\textrm{YM}}\left[\tilde{A}\right]} \int \mathcal{D} \delta A \, \exp\left(-\frac{1}{4g^2}\left.\frac{\delta^2 S_{\textrm{YM}}}{\delta A^2}\right|_{\tilde A} \delta A^2\right)
\end{split}
\end{equation}
Hence the source term is cancelled by the equation of motion for $\tilde A$ and by the Legendre transform, and therefore it does not contribute to the \emph{$1$PI} effective action. \par
For completeness we recall the connection between the aforementioned standard definition of the generating functional and the gauge invariant \emph{$1$PI} effective action in the background field method~\cite{Abbott:1981ke}. To define a gauge invariant effective action, we introduce a modified
generating functional depending on the background field:
\begin{equation}
Z[J,\tilde A] = \int \mathcal{D}\delta A\,\exp\left[-\frac{1}{2g^2}S_{\textrm{YM}}(\tilde A+\delta A) + J \delta A\right]
\end{equation}
where we split the gauge field $A=\tilde A+\delta A$ into a background field $\tilde A$ and a quantum field $\delta A$. We omit for the moment the gauge fixing condition for the quantum field.
If we assume that the source transforms as:
\begin{subequations}
\begin{align}
\delta_c J_m&= i \left[J_m,\Theta\right]
\intertext{the \emph{YM} action plus the source term are invariant for the classical symmetry:}
\delta_c \tilde A_m &= D_m(\tilde A) \,\Theta \\
\delta_c \delta A_m &= i \left[\delta A_m,\Theta\right]
\intertext{and for the quantum symmetry:}
\delta_q \tilde A_m &= 0\\
\delta_q \delta A_m &= D_m(\delta A) \,\Omega +i [\tilde A_m,\Omega]
\end{align}
\end{subequations}
Therefore, if we choose a gauge-fixing condition for the quantum field invariant for the classical symmetry, e.g.:
\begin{equation}
\mathcal{G}=D_m(\tilde A)\delta A_m-c
\end{equation}
the \emph{YM} action, the Fadeev-Popov (\emph{FP}) determinant and the modified generating functional are invariant for the classical symmetry as well.
The same classical symmetry holds for the connected generating functional and for the effective action:
\begin{subequations}
\begin{align}
G[J,\tilde {A}] &= - \log(Z[J,\tilde A])\\
\Gamma[\widetilde{\delta A},\tilde{A}] &= G [J,\tilde A] - J \widetilde{\delta A}
\end{align}
where:
\begin{equation}
\begin{aligned}
\widetilde{\delta A} &= \frac{\delta G[J,\tilde A]}{\delta J},& \delta_c \widetilde{\delta A} = i[\widetilde{\delta A},\Theta]
\end{aligned}
\end{equation}
\end{subequations}
In order to relate this formalism to the standard one, we perform the change of variables $\delta A \to \delta A - \tilde A$ in the functional integral. The classical field disappears from the \emph{YM} action, but it is still present in the gauge fixing:
\begin{equation}
Z[J,\tilde A] = \exp(-J \tilde A) \int \mathcal{D}\delta A \exp\left[-\frac{1}{2g^2}S_\text{YM}\left(\delta A\right)+ J \delta A\right] \equiv \exp(- J \tilde A) \,\tilde Z_{\tilde A}[J]
\end{equation}
in such a way that $\tilde Z_{\tilde A}[J]$ is the standard generating functional but with the unusual gauge-fixing:
\begin{equation}
\mathcal{G}[\delta A] = D_m (\tilde A)\left(\delta A_m - \tilde A_m\right)-c
\end{equation}
that reduces to:
\begin{subequations}
\begin{equation}
\mathcal{G}[\delta A] = D_m (\tilde A)\, \delta A_m -c
\end{equation}
provided:
\begin{equation}\label{eq:backgroundgauge}
\partial_m \tilde A_m=0
\end{equation}
\end{subequations}
The $\tilde Z$ generating functional leads to the connected generating functional:
\begin{align}
G[J,\tilde A] &= J \tilde A + \tilde G_{\tilde A}[J]\\
\intertext{and to the effective action:}
\Gamma[\widetilde{\delta A},\tilde A] &= \tilde\Gamma_{\tilde A}[\widetilde{\delta A}+\tilde A]
\end{align}
Finally, we can set $\widetilde{\delta A} = 0$ in such a way that the gauge invariant effective action coincides with the standard one in the aforementioned unusual gauge. \par
For the reader convenience we report the result of the standard computation of the one-loop effective action in the Feynman gauge, since it is needed to prove the identity with the one-loop effective action in the $ASD$ variables (for a detailed derivation see Section 3.1 of ref. \cite{Bochicchio:2011en}). \par
After integrating on the quadratic fluctuation $\delta A$ and inserting the \emph{FP} determinant $\Det(- \Delta)$, the one-loop effective action $\Gamma_\textrm{1-loop}({A})$ reads:
\begin{equation}
\label{eq:Z1loop}
e^{-\Gamma_\textrm{1-loop}({A})}
=e^{-\frac{1}{2g^2}S_\text{YM} \left({A}\right)}{\Det}^{-1/2}\left(-\Delta \delta_{m n}-2i \ad F_{m n}\right) \Det(-\Delta)
\end{equation}
where we omit the superscript $\tilde \,$ over background fields, and we define $\ad F_{m n}=[ F_{m n},\cdot]$, i.e.:
\begin{equation} 
\label{ad}
(\ad F_{m n})^{ac} = if^{abc} F_{m n}^b
\end{equation}
Using the identity:
\begin{equation}
\Det^{-1/2}\left(-\Delta \,\delta_{m n}-2i \ad F_{m n}\right)= \Det^{-1/2}\left(-\Delta\, \delta_{m n}\right)\Det^{-1/2}\left(1-2i\left(-\Delta\right)^{-1} \, \ad F_{m n}\right)
\end{equation}
the first factor gives:
\begin{equation}
\Det^{-1/2}\left(-\Delta\,\delta_{m n}\right)=\Det^{-2}\left(-\Delta\right)
\end{equation}
Therefore, the one-loop effective action reads:
\begin{equation}
\label{eq:orbitalspin}
e^{-\Gamma_\text{1-loop}\left({A}\right)}=e^{-\frac{1}{2g^2}S_\text{YM} \left(A\right)}\Det^{-1/2}\left(1-2i\left(-\Delta\right)^{-1} \, \ad F_{m n}\right) \Det^{-1}\left(-\Delta\right)
\end{equation}
The first determinant is the spin contribution, the second determinant is the orbital contribution.
From this expression we extract the local part of the one-loop effective action:
\begin{equation}
\label{eq:standard}
\begin{split}
\Gamma_\textrm{1-loop}& \sim \frac{1}{2g^2}S_\text{YM}+\left(\frac{N}{3\left(4\pi\right)^2}-\frac{4N}{\left(4\pi\right)^2}\right) \log\left(\frac{\Lambda}{M}\right) \frac{1}{2}\int d^4 x \left(F^a_{m n}\right)^2 \\
&=\left(\frac{1}{2g^2\left(\Lambda\right)}-\frac{11N}{3\left(4\pi\right)^2}\log\left(\frac{\Lambda}{M}\right)\right) S_\text{YM}
\end{split}
\end{equation}
where $\Lambda$ is the ultraviolet cutoff and $M$ an infrared scale.
The first term in front of the logarithm in the first line is the orbital contribution to the one-loop beta function of the $YM$ theory, the second term is the spin contribution.

\section{\texorpdfstring{Identity of the one-loop \emph{$1$PI} effective actions in the covariant \mbox{$\alpha$-gauge}}{Identity of the one-loop 1PI effective actions in the covariant alpha-gauge}}
\label{sec:effective}

We show now that the Jacobian and the extra determinant due to the external source in the \emph{ASD} variables in the covariant $\alpha$-gauge combine to produce the determinant obtained integrating over the gauge connection
in the original variables. \par
Therefore, in the $\alpha$-gauge the effective action in the \emph{ASD} variables is identical to the one in the original variables. \par
We express the term $J A$ as a function of $\nu$, first computing $J$ as a function of $\tilde\nu$ at the saddle point and then inverting perturbatively the map from $A$ to $\nu$.
We simplify the notation omitting the superscript $^-$ over $\mu$ and $\nu$. \par
The gauge-fixed action density with the source term is:
\begin{equation}
\frac{1}{4g^2} {\nu^{\dot \alpha}}_{\dot \beta} \overalpha {\nu_{\dot \alpha}}^{\dot \beta} - \bar J^{\dot \alpha \beta} A_{\beta \dot \alpha }
\end{equation}
with ${\nu^{\dot \alpha}}_{\dot \beta} \overalpha = {\nu^{\dot \gamma}}_{\dot \delta}\left[\delta_{\dot\gamma}^{\dot\alpha}\delta_{\dot\beta}^{\dot\delta} - \frac12\left(1-\frac1{\alpha}\right)\delta_{\dot\beta}^{\dot\alpha}\delta_{\dot\delta}^{\dot\gamma}\right]$. The saddle point equation, i.e. the equation of motion with the source term, is:
\begin{equation}
\begin{split}
\frac{1}{2g^2} \left.{\tilde{\nu}^{\dot \alpha}}\right._{\dot \beta} \overalpha &= \bar J^{\dot \gamma \delta} \left.\frac{\delta A_{\delta \dot \gamma }}{\delta{\nu_{\dot \alpha}}^{\dot \beta}}\right|_{\tilde\nu}\\
\bar J^{\dot \gamma \delta} &= \frac{1}{2g^2} \left.\tilde{\nu}^{\dot \alpha}\right._{\dot \beta} \overalpha \left.\frac{\delta{\nu^{\dot \beta}}_{\dot \alpha}}{\delta A_{\delta \dot \gamma }}\right|_{A\left(\tilde\nu\right)}\\
\bar J^{\dot \gamma \delta}&= - \frac{1}{2g^2} \left.\tilde{\nu}^{\dot \gamma}\right._{\dot \beta} \overalpha \overleftarrow{\bar{D}}^{\dot \beta \delta}
\end{split}
\end{equation}
We can choose the source term in such a way that $c=0$ at the saddle point, i.e. in such a way that $\tilde\nu = \tilde\mu$, with $\mu$ the anti-Hermitian traceless matrix defined in Eq.~\eqref{eq:mumenospinor}. This choice of 
gauge for the background field at the saddle point ensures the gauge invariance of the effective action according to Eq.~\eqref{eq:backgroundgauge}.
Thus the equation of motion reduces to:
\begin{equation}
\bar J^{\dot \gamma \delta}= - \frac{1}{2g^2} \left.\tilde\mu^{\dot \gamma}\right._{\dot \beta} \overleftarrow{\bar{D}}^{\dot \beta \delta}
\end{equation}
Nevertheless, the fluctuation of $c$ does not vanish, and therefore we maintain the distinction between the fluctuations $\delta \nu$ and $\delta \mu$ in the following.
To find $A$ as a function of $\nu$ we expand $F^-$ around the background field $\tilde A$:
\begin{equation}
\begin{split}
F(\tilde A + \delta A)^a_{mn} &= F(\tilde A)^a_{mn} + D(\tilde A)^{ac}_m \delta A^c_n - D(\tilde A)^{ac}_n \delta A^c_m - f^{abc}\delta A^b_m \delta A^c_n \\
F^-(\tilde A + \delta A)^a_{mn} &= F^-(\tilde A)^a_{mn} + D(\tilde A)^{ac}_m \delta A^c_n - D(\tilde A)^{ac}_n \delta A^c_m - \epsilon_{mnrs} D(\tilde A)^{ac}_r \delta A^c_s\\
&\quad - f^{abc}\left(\delta A^b_m \delta A^c_n - \frac12 \epsilon_{mnrs} \delta A^b_m \delta A^c_n\right)
\end{split}
\end{equation}
where $D(\tilde A)$ is the covariant derivative with respect to the background field.
We split $\nu = \tilde\nu + \delta\nu$ with $\tilde\nu = \tilde\mu = F^-(\tilde A)$ the background field in the \emph{ASD} variables. From now on, to simplify the notation, we omit the $\,\tilde\,\,$ over all the background fields since no confusion can arise. Moreover, the dependence of the covariant derivatives on the background field is understood:
\begin{equation}
\delta\nu^a_{mn} = D^{ac}_m \delta A^c_n - D^{ac}_n \delta A^c_m - \epsilon_{mnrs} D^{ac}_r \delta A^c_s - f^{abc}\left(\delta A^b_m \delta A^c_n - \frac12 \epsilon_{mnrs} \delta A^b_m \delta A^c_n\right)
\end{equation}
that in the spinor-matrix representation reads:
\begin{equation}
\delta\nu^a = \bar{D}^{ac} \delta A^c - \frac12 f^{abc} \overline{\delta A}^b \delta A^c
\end{equation}
as we found in Eq.~\eqref{eq:mapnu} by using the spinor notation from the very start. This equation can be inverted around the background field perturbatively in the fluctuation. To the order quadratic in the fluctuation, necessary for the one-loop computation, we get:
\begin{equation}\label{eq:adimu}
\begin{split}
- D^{da}\bar{D}^{ac} \delta A^c &= -D^{da} \delta\nu^a - \frac12 f^{abc} D^{da} \left(\overline{\delta A}^b \delta A^c\right)\\
\delta A^e &= -G^{ed} D^{da} \delta\nu^a - \frac12 f^{abc} G^{ed} D^{da} \left(\overline{\delta A}^b \delta A^c\right)\\
\end{split}
\end{equation}
where $G^{ed} = \left[\left(-D \bar D \right)^{-1}\right]^{ed}$. 

The first term in Eq.~\eqref{eq:adimu} is linear in the fluctuation $\delta\nu$ and therefore it is irrelevant for the effective action, as it is the source term in the original variables.
The second term in Eq.~\eqref{eq:adimu} is quadratic in the fluctuation $\delta\nu$. We keep it and we get:

\begin{equation}
\begin{split}
\left(\tr \bar J^a \delta A^a\right)^{(2)} = \frac{1}{4g^2} f^{edg} \tr  \left[ \mu^f (\overleftarrow{\bar{D}})^{fa} \left(G D\right)^{ae} \left(\delta\nu^b (\overleftarrow{G\bar{D}})^{bd} \left(G D\right)^{gc} \delta\nu^c\right) \right]
\end{split}
\end{equation}
Integrating by parts, it simplifies to:
\begin{equation}
\begin{split}
\left(\tr \bar J^a \delta A^a\right)^{(2)} = -\frac{1}{4g^2} f^{adg} \tr \left[ \mu^a \delta\nu^b \left(\bar{D} G\right)^{bd} \left(G D\right)^{gc} \delta\nu^c \right]
\end{split}
\end{equation}
Explicitly, it reads:
\begin{equation}
\label{color}
\begin{split}
\left(\tr \bar J^a \delta A^a\right)^{(2)} &= \frac{f^{adg}}{4g^2}\!\int\!d^4x \,d^4z\,d^4t\, {\left(\mu^a_x\right)^{\dot \alpha}}_{\dot \beta} {(\delta\nu^b_z)^{\dot \beta}}_{\dot \gamma} (\bar{D}_z^{be})^{\dot \gamma \delta} (G_{xz}^{ed})_{\delta}^{\phi} (G_{xt}^{gf})_{\phi}^{\lambda} (D_t^{fc})_{\lambda \dot \rho} {(\delta\nu^c_t)^{\dot \rho}}_{\dot\alpha}\\
& \equiv \frac{1}{4g^2} \int d^4z\,d^4t\, {(\delta\nu^b_z)^{\dot \beta}}_{\dot \gamma} \left(O^{bc}_{zt}\right)_{\dot\beta\dot\rho}^{\dot\gamma\dot\alpha} {(\delta\nu^c_t)^{\dot \rho}}_{\dot\alpha}
\end{split}
\end{equation}
where, to keep the notation compact, we write the space-time coordinates of fields as indices, e. g. $\mu_x \equiv \mu(x)$, we add a superscript to the covariant derivatives to specify on which space-time point they act, we set $G_{xt} \equiv G\left(x-t\right)$, and similarly we denote the space-time delta function here below by $\delta_{zt}$.
We now integrate over the fluctuation $\delta \nu$ to get:
\begin{align} \label{eq:sourcedet}
&\int \mathcal{D}\delta \nu \Det\frac{\delta A}{\delta\nu} \exp\left[-\frac{1}{2g^2}S_{\textrm{YM}} (\mu + \delta \nu) + JA( \mu+ \delta \nu) \right]  \nonumber \\
& =\exp\left[-\frac{1}{2g^2}S_{\textrm{YM}}(\mu)+ JA( \mu)\right] \Det\frac{\delta A}{\delta\nu}  \caldet{J}  \nonumber \\
\end{align}
where we have defined:
\begin{equation}
 \caldet{J} \equiv \Det^{-1/2}\left[- \left(\overalpha\right)_{\dot\beta\dot\rho}^{\dot\gamma\dot\alpha} \delta^{bc} \delta_{zt} + \left[O^{bc}_{zt}\right]_{\dot\beta\dot\rho}^{\dot\gamma\dot\alpha}\right]
\end{equation}
Finally, we evaluate the Jacobian of the map to the \emph{ASD} variables. From the definition Eq.~\eqref{eq:mapnu} we get:
\begin{equation}
\frac{\delta A}{\delta\nu} = \bar D^{-1}
\end{equation}
The infinite determinant of a non-Hermitian operator is ill-defined. It can be given a meaning by multiplying by the adjoint and taking the square root:
\begin{equation}
\Det\frac{\delta A}{\delta\nu} = \Det^{-1/2}\left(D \bar D\right)
\label{eq:jacobian}
\end{equation}
Therefore, the product of the determinants in Eq.~\eqref{eq:sourcedet} reads:
\begin{equation}
\begin{split}
\caldet{\nu}& \equiv \caldet{J}\Det\frac{\delta A}{\delta \nu}= \Det^{-1/2}\left(D\right) \Det^{-1/2}\left[-\overalpha + O \right]\Det^{-1/2}\left(\bar D\right)\\
&=\Det^{-1/2}\left[-D\overalpha \bar D + D O \bar D\right]
\end{split}
\end{equation}
and working out in detail the spinor notation:
\begin{equation}
\caldet{\nu} =\Det^{-1/2}\left[- \left(D\bar D\right)_\gamma^\rho \delta^{\dot\alpha}_{\dot\beta} + \frac12\left(1-\frac{1}{\alpha}\right) D_{\gamma\dot\beta} \bar{D}^{\dot\alpha\rho} + D^z_{\gamma\dot\gamma} \left[O_{zt}\right]_{\dot\beta\dot\rho}^{\dot\gamma\dot\alpha} (\bar D^{t})^{\dot\rho\rho} \right]
\end{equation}
The operator $D \bar D$ can be evaluated as:
\begin{equation} \label{33}
\begin{split}
D\bar D &= D_m \sigma^m D_n \bar \sigma^n = \frac12 D_m D_n \left(\sigma^{\{m}\bar \sigma^{n\}} + \sigma^{[m}\bar \sigma^{n]}\right)\\
&= D_m D_n \delta^{mn}\mathds{1} + i \frac{1}{4} \ad F(\tilde A)_{mn} \sigma^{[m}\bar \sigma^{n]} = \Delta \mathds{1} + i \frac12 \ad F^+(\tilde A)_{mn} \sigma^{mn} \\
&= \Delta \mathds{1} + i \ad F(\tilde A)^+
\end{split}
\end{equation}
with ${(F(\tilde A)^+)_{\alpha}}^{\beta} = \frac12 F(\tilde A)^+_{mn} {\left(\sigma^{mn} \right)_{\alpha}}^{\beta}$ and $(\ad F)$ defined in Eq.~\eqref{ad}. To simplify the notation the dependence of $F$ on the background field is understood. Hence using the previous equation for $D\bar D$ and the action of the covariant derivatives on $O$ we get:
\begin{equation}
\caldet{\nu} =\Det^{-1/2}\left[- \Delta \delta_\gamma^\rho \delta^{\dot\alpha}_{\dot\beta} -i {(\ad F^{+})_{\gamma}}^{\rho} \delta^{\dot\alpha}_{\dot\beta} + \frac12\left(1-\frac{1}{\alpha}\right) D_{\gamma\dot\beta} \bar{D}^{\dot\alpha\rho}- i {\left(\ad \mu\right)^{\dot\alpha}}_{\dot\beta} \delta_\gamma^\rho \right]
\end{equation}
Going back to the vector notation and using $F^+ + \mu = F^+ + F^- = 2F$ we obtain:
\begin{equation}
\begin{split}
\caldet{\nu}&=\Det^{-1/2}\left[- \Delta \delta_{mn} + \left(1-\frac{1}{\alpha}\right) D_m D_{n} - i \ad F^{+}_{mn} - i \ad \mu_{mn} \right]\\
&= \Det^{-1/2}\left[- \Delta \delta_{mn} + \left(1-\frac{1}{\alpha}\right) D_m D_{n} - 2 i \ad F_{mn}\right] \\
& \equiv \caldet{A} \\
\end{split}
\end{equation}
This is precisely the determinant associated to the fluctuation of the gauge connection in the original variables in the $\alpha$-gauge. Together with the \emph{FP} determinant it furnishes the one-loop quantum contribution to the \emph{$1$PI} effective action, that for $\alpha=1$ has been computed in Eq.~\eqref{eq:Z1loop}.

\section{\texorpdfstring{Identity of the one-loop \emph{$1$PI} effective actions in an arbitrary gauge}{Identity of the one-loop 1PI effective actions in an arbitrary gauge}}
\label{sec:geometric}
We introduce now a geometric notation. We define the 1-forms:
\begin{subequations}
\begin{align}
A &= A_m\,dx_m & J &= J_m\,dx_m
\intertext{and the 2-forms:}
F &= F_{mn}\,dx_m \wedge dx_n & \mu &= \mu^-_{mn} \,dx_m \wedge dx_n
\end{align}
where:
\begin{equation}
F = d A + i A \wedge A
\end{equation}
\end{subequations}
The projectors over the \emph{SD} (resp. \emph{ASD}) curvatures are:
\begin{equation} \label{5}
P^\pm_{mn,rs} = \frac14 \left(\delta_{mr} \delta_{ns} - \delta_{ms} \delta_{nr} \pm \epsilon_{mnrs}\right)
\end{equation}
In Euclidean space-time the projectors are eigenstates of the Hodge star:
\begin{equation}
{*P}^\pm = \pm {P^\pm}
\end{equation}
with $** = 1$. With this notation, omitting the trace on color indices everywhere in the following, the generating functional reads:
\begin{equation}
Z[J] = \int \mathcal{D}\mu\,\exp\left[-\frac{1}{8g^2} \int \mu \wedge * \mu + \int *J \wedge A\right] \delta\left(\mathcal{G}\left(A\right)\right) \Det\frac{\delta A}{\delta \mu} \, \Delta_\text{FP}
\end{equation}
and since $\mu$ is \emph{ASD}:
\begin{equation}
Z[J] = \int \mathcal{D}\mu\,\exp\left[\frac{1}{8g^2} \int \mu \wedge \mu + \int *J \wedge A\right] \delta\left(\mathcal{G}\left(A\right)\right) \Det\frac{\delta A}{\delta \mu} \, \Delta_\text{FP}
\end{equation}
where $\mathcal{G}$ is the gauge-fixing functional and $\Delta_\text{FP}$ the corresponding \emph{FP} determinant. The change to the \emph{ASD} variables is the map:
\begin{subequations}
\begin{equation} \label{34}
\mu = 2 P^- F = 2 P^- \left(F(\tilde A) + d_{\tilde A} \wedge \delta A + i \delta A \wedge \delta A\right)
\end{equation}
where $d_{\tilde A}$ is the covariant derivative along the classical field $\tilde A$. In general gauges $\mathcal{G}\left[A\right]=0$ we simply restrict this map to the gauge-fixed slice of the gauge orbits,
in such a way that its linearization defines an invertible operator in perturbation theory.
We split $\mu$ into a background field $\tilde \mu$, depending only on the background gauge connection $\tilde A$, and a quantum fluctuation $\delta\mu$:
\begin{align}
\tilde \mu &= 2P^- F(\tilde A) \\
\delta \mu &= 2P^- \left(d_{\tilde A} \wedge \delta A + i \delta A \wedge \delta A\right) \label{eq:geommapping}
\end{align}
\end{subequations}
so that $\mu = \tilde \mu(\tilde A) + \delta\mu(\tilde A, \delta A)$. In the following we omit the $\,\tilde\,\,$ symbol over classical fields since no confusion can arise. Once restricted to the gauge-fixed slice Eq.~\eqref{eq:geommapping} can be inverted perturbatively:
\begin{subequations}
\begin{align}
\delta A &= \delta A^{(1)} + \delta A^{(2)} + \dots \\
\delta A^{(1)} &= \frac12 \left(P^- d_A \wedge \right)^{-1} \delta \mu\label{eq:mapping1}\\
\delta A^{(2)} &= -\frac{i}4 \left(P^- d_A \wedge \right)^{-1} \left[\left(P^- d_A \wedge \right)^{-1}\delta \mu \wedge \left(P^- d_A \wedge \right)^{-1}\delta \mu\right]\label{eq:mapping2}
\end{align}
\end{subequations}
The equation of motion reads:
\begin{equation}
\mu = -4g^2 * J \wedge \left.\frac{\delta A}{\delta\mu}\right|_{\mu}
\end{equation}
From Eq.~\eqref{eq:mapping1} we evaluate:
\begin{equation}
\label{2}
\frac{\delta A}{\delta\mu} = \frac12\left(P^- d_A \wedge\right)^{-1}
\end{equation}
so that:
\begin{align}
\mu &= 2g^2 \left(P^- d_A\wedge\right)^{-1} \wedge *J\\
J &= \frac{1}{2g^2} *\left(P^- d_A\wedge \mu\right)
\end{align}
$\delta A^{(2)}$ and the saddle point equation determine the quadratic form in the fluctuation that contributes to the effective action:
\begin{equation}
\exp\left[\frac{1}{8g^2} \int \mu \wedge \mu + \int *J \wedge A\right]\sim \exp\left[\frac{1}{8g^2} \int \delta\mu \wedge \delta\mu +\int *J \wedge \frac{1}{2}  \frac{\delta^2 A}{\delta\mu^2} \delta\mu^2\right]
\end{equation}
The source term reads:
\begin{equation}
\int *J \wedge \frac{1}{2} \frac{\delta^2 A}{\delta\mu^2} \delta\mu^2 = \int \frac{1}{2g^2} \left(P^- d_A\wedge \mu\right) \wedge \delta A^{(2)}
\end{equation}
with:
\begin{equation}
\begin{split}
\int \frac{1}{2} \left(P^- d_A\wedge \mu\right) \wedge \delta A^{(2)} & = \int
-\frac{i}{8} \left(P^- d_A \wedge \mu\right) \wedge \left(P^- d_A \wedge \right)^{-1} \left[\left(P^- d_A \wedge \right)^{-1}\delta \mu \wedge \left(P^- d_A \wedge \right)^{-1}\delta \mu\right]\\
&= \int \frac{i}{8} \mu \wedge\left(P^- d_A \wedge \right)^{-1}\delta \mu \wedge \left(P^- d_A \wedge \right)^{-1}\delta \mu\\
&= \int \frac{i}{8} 2 P^- F \wedge\left(P^- d_A \wedge \right)^{-1}\delta \mu \wedge \left(P^- d_A \wedge \right)^{-1}\delta \mu
\end{split}
\end{equation}
where we have integrated by parts and substituted $ \mu = 2 P^- F$.
Hence the quadratic form is:
\begin{multline}
\frac{1}{8g^2} \int \delta\mu \wedge \delta\mu + \frac{1}{2} \int *J \wedge \frac{\delta^2 A}{\delta\mu^2} \delta\mu^2 \\= \frac{1}{8g^2} \Bigg[\int \delta\mu \wedge \delta\mu + 2i \int P^- F \wedge\left(P^- d_A \wedge \right)^{-1}\delta \mu \wedge \left(P^- d_A \wedge \right)^{-1}\delta \mu \Bigg]
\end{multline}
The Jacobian to the \emph{ASD} variables reads in geometric notation:
\begin{equation}
\Det \frac{\delta A}{\delta \mu} = \Det \left(P^- d_A \wedge \right)^{-1} = \Det^{-1/2} \left(P^- d_A \wedge \right) \Det^{-1/2} \left(P^- d_A \wedge \right)
\end{equation}
where we have dropped an irrelevant factor of $\frac{1}{2}$ from Eq.~\eqref{2} in the functional derivative.
Combining the two factors with the determinant due to the quadratic fluctuation in the source term we get:
\begin{multline}
\int \mathcal{D}\mu \,\exp\left[\frac{1}{8g^2} \int \delta\mu \wedge \delta\mu +\frac{1}{2} \int *J \wedge \frac{\delta^2 A}{\delta\mu^2} \delta\mu^2\right] \Det \frac{\delta A}{\delta\mu} \\
= \Det^{-1/2} \left(P^- d_A \wedge \right) \Det^{-1/2} \left[1 + 2i \left(P^- d_A \wedge \right)^{-1} P^- F \wedge \left(P^- d_A \wedge \right)^{-1} \right] \Det^{-1/2} \left(P^- d_A \wedge \right)\\
=\Det^{-1/2} \left[ P^- d_A \wedge P^- d_A \wedge + 2i P^- F \wedge \right]\label{eq:detmapping}
\end{multline}
This has to be compared with the effective action in geometric notation in the original variables. The partition function reads:
\begin{equation}
Z = \int \mathcal{D}A\, \exp \left[\frac{1}{8g^2} \int F^- \wedge F^-\right] \Delta_\text{FP} \,\delta\left(\mathcal{G}\left(A\right)\right)
\end{equation}
Expanding $F^-$ around $\tilde A$ by means of Eq.~\eqref{34}, the quadratic form in the fluctuation $\delta A$ turns out to be ~\cite{Bochicchio:2011en}:
\begin{equation}
\left(P^- d_A \wedge \delta A \right) \wedge \left(P^- d_A \wedge \delta A \right) + 2i P^- F \wedge \delta A \wedge \delta A
\end{equation} 
up to a multiplicative constant.
The associated determinant restricted to the gauge-fixed slice is:
\begin{equation}
\caldet{A}=\Det^{-1/2} \left[ P^- d_A \wedge P^- d_A \wedge + 2i P^- F \wedge \right]
\end{equation}
that coincides with Eq.~\eqref{eq:detmapping}.

\section{One-loop beta function in the \emph{ASD} variables in the Feynman gauge}
\label{sec:feynman}
To get the one-loop beta function directly in the \emph{ASD} variables, we evaluate the divergent part of the determinant in Eq.~\eqref{eq:sourcedet} in the Feynman gauge $\alpha=1$. With the accuracy necessary for the one-loop computation of the beta function we can substitute to all the covariant derivatives the ordinary derivatives in the operator $O$. Indeed, we are looking for a counterterm proportional to $\int \tr(\mu^2) d^4x$. This counterterm may arise in the functional determinant by a term at most quadratic in $O$, since $O$ is at least linear in $\mu=F^-$. \par
The term linear in $O$ could in principle contribute, but in fact it does not. Indeed, expanding $G$ in powers of $F^+$ by means of Eq.~\eqref{33}, 
the zero-order term contains the inverse of the covariant Laplacian that is a spin singlet, and thus it vanishes because of the trace on spin and the insertion of $\mu$, that is traceless. Besides, the counterterm  linear in $F^+$ is $\tr(F^+ F^-)$, that vanishes identically. The other terms are irrelevant at this order. \par
Moreover, the counterterm quadratic in $O$ is already
quadratic in $\mu$, and thus the extra insertions of the background field, due to the covariant derivatives and to $F^+$ in the expansion of $G$, contribute only higher-order terms in $\mu$. Therefore, with the necessary accuracy:
\begin{equation}
\begin{split} 
&\frac{1}{4g^2} \int d^4z\,d^4t\, {(\delta\nu^b_z)^{\dot \beta}}_{\dot \gamma} \left(O^{bc}_{zt}\right)_{\dot\beta\dot\rho}^{\dot\gamma\dot\alpha} {(\delta\nu^c_t)^{\dot \rho}}_{\dot\alpha} \\
&\sim \frac{f^{abc}}{4g^2}\!\int\!d^4x \,d^4z\,d^4t\, {\left(\mu^a_x\right)^{\dot \alpha}}_{\dot \beta} {(\delta\nu^b_z)^{\dot \beta}}_{\dot \gamma} (\bar{\partial}^z)^{\dot \gamma \delta} G_0\left(x-z\right) G_0\left(x-t\right) \left(\partial^t\right)_{\delta \dot \rho} {(\delta\nu^c_t)^{\dot \rho}}_{\dot\alpha}\\
\end{split}
\end{equation}
where $G_0 = \left(-\partial^2\right)^{-1}$. We have:
\begin{equation}
\begin{split}
\caldet{J}&=\Det^{-1/2}\left[- \delta^{\dot\alpha}_{\dot\beta}\delta_{\dot\rho}^{\dot\gamma} \delta^{bc} \delta_{zt} + \left[O^{bc}_{zt}\right]_{\dot\beta\dot\rho}^{\dot\gamma\dot\alpha}\right]\\
&\sim\exp\left[\frac{1}4 \int d^4t \,d^4z \, \left[O^{bc}_{zt}\right]_{\dot\beta\dot\rho}^{\dot\gamma\dot\alpha} \delta^{\dot\rho}_{ \dot \gamma'} \delta_{\dot \alpha}^{\dot \beta'}\left[O^{cb}_{tz}\right]_{\dot\beta'\dot\rho'}^{\dot\gamma'\dot\alpha'} \delta^{\dot\rho'}_{ \dot \gamma} \delta_{\dot \alpha'}^{\dot \beta}\right]
\end{split}
\end{equation}
The integral reads:
\begin{equation}
\begin{split}
\log\caldet{J}&\sim\frac14 f^{abc} f^{a'cb} \int d^4x\, d^4 x' \,d^4z\,d^4t\quad {\Big(\mu^a_x\Big)^{\dot \alpha}}_{\dot \beta} \left({\bar\partial}_z^{\dot \gamma \delta} G_0\left(x-z\right) G_0\left(x-t\right) \partial^t_{\delta \dot \rho} \right)\\
&\qquad\delta^{\dot\rho}_{ \dot \gamma'} \delta_{\dot \alpha}^{\dot \beta'} \delta^{\dot\rho'}_{ \dot \gamma} \delta_{\dot \alpha'}^{\dot \beta}\quad{\left(\mu^{a'}_{x'}\right)^{\dot \alpha'}}_{\dot \beta'} \left({\bar\partial}_t^{\dot \gamma' \delta'} G_0\left(x-t\right) G_0\left(x-z\right) \partial^z_{\delta' \dot \rho'} \right)
\end{split}
\end{equation}
Contracting the Lie algebra indices, we get $f^{abc} f^{a'cb} = - N \delta^{aa'}$. Taking into account this factor of $N$, we suppress in the following the Lie algebra indices of the field $\mu$. In momentum space we get:
\begin{multline}
\log\caldet{J}\sim-\frac14 N \int d^4k\, d^4p \frac{1}{\left(2\pi\right)^{8}} \,{\left(\mu_p\right)^{\dot \alpha}}_{\dot \beta} {\left(\mu_{-p}\right)^{\dot \alpha'}}_{\dot \beta'} \\\left(\bar{\slashed{k}}^{\dot \gamma \delta}\slashed{k}_{\delta' \dot \rho'} \left(\slashed p +\slashed{k}\right)_{\delta \dot \rho} \left(\bar{\slashed p} +\bar{\slashed{k}}\right)^{\dot \gamma' \delta'}\right)
\delta^{\dot\rho}_{ \dot \gamma'} \delta_{\dot \alpha}^{\dot \beta'} \delta^{\dot\rho'}_{ \dot \gamma} \delta_{\dot \alpha'}^{\dot \beta}\, \frac{1}{k^4 \left(p+k\right)^4}
\end{multline}
We are interested in the logarithmic divergences. All the terms $O\left(\frac{p}{k}\right)$ are finite, therefore we can discard them:
\begin{equation}
\begin{split}
\log\caldet{J}&\sim-\frac14 N \int d^4k\, d^4p \frac{1}{\left(2\pi\right)^{8}} \, {\left(\mu_p\right)^{\dot \alpha}}_{\dot \beta} {\left(\mu_{-p}\right)^{\dot \alpha'}}_{\dot \beta'} \left(\bar{\slashed{k}}^{\dot \gamma \delta}\slashed{k}_{\delta' \dot \rho'} \slashed{k}_{\delta \dot \rho} \bar{\slashed{k}}^{\dot \gamma' \delta'}\right)
\delta^{\dot\rho}_{ \dot \gamma'} \delta_{\dot \alpha}^{\dot \beta'} \delta^{\dot\rho'}_{ \dot \gamma} \delta_{\dot \alpha'}^{\dot \beta} \frac{1}{k^8}\\
\quad&= -\frac14 N \int\, d^4k \, d^4p \frac{1}{\left(2\pi\right)^{8}} \,{\left(\mu_p\right)^{\dot \alpha}}_{\dot \beta} {\left(\mu_{-p}\right)^{\dot \alpha'}}_{\dot \beta'} \left(\delta^{\dot \gamma}_{\dot \rho} \delta_{\dot\rho'}^{\dot\gamma'}\right)\,\delta^{\dot\rho}_{ \dot \gamma'} \delta_{\dot \alpha}^{\dot \beta'} \delta^{\dot\rho'}_{ \dot \gamma} \delta_{\dot \alpha'}^{\dot \beta} \frac{1}{k^4}\\
&=- \frac14 2N \int \frac{d^4p}{\left(2\pi\right)^4} \,{\left(\mu_p\right)^{\dot \alpha}}_{\dot \beta} {\left(\mu_{-p}\right)^{\dot \beta}}_{\dot \alpha} \int \frac{d^4k}{\left(2\pi\right)^{4}k^4}\\
&= - \frac{2N}{\left(4\pi\right)^2} \log\left(\frac{\Lambda}{M}\right) \frac12\int d^4x \,\, {\left(\mu \right)^{\dot \alpha}}_{\dot \beta} {\left(\mu \right)^{\dot \beta}}_{\dot \alpha}
\end{split}
\end{equation}
${\left(\mu \right)^{\dot \alpha}}_{\dot \beta} {\left(\mu \right)^{\dot \beta}}_{\dot \alpha}$ is negative definite because the matrices $\mu$ are anti-Hermitian and thus a sign changes with respect to the action. The divergent local part of the determinant reads:
\begin{equation}
\label{eq:detsource}
\begin{split}
\caldet{J}&\sim \exp\left[- \frac{2N}{\left(4\pi\right)^2} \log\left(\frac{\Lambda}{M}\right)\,\frac{1}{2} \int d^4x \,{\left(\mu^a \right)^{\dot \alpha}}_{\dot \beta} {\left(\mu^a \right)^{\dot \beta}}_{\dot \alpha}\right]\\
&= \exp\left[\frac{2N}{\left(4\pi\right)^2} \log\left(\frac{\Lambda}{M}\right)\, 2   \int d^4x\, \tr (P^- F)_{mn}^2   \right]
\end{split}
\end{equation}
We combine the divergent part of this determinant with those of the Jacobian and of the \emph{FP} determinant. The Jacobian decomposes into an orbital contribution and a spin contribution (see for comparison Eq.~\eqref{eq:orbitalspin} and following):
\begin{equation}
\begin{split}
\Det \frac{\delta A}{\delta \nu}&=\Det^{-1/2} \left(D\bar D\right)= \Det^{-1/2} \left[-\Delta\delta_{mn} - i\ad F^+_{mn} \right] \\&= \Det^{-1/2} \left[-\Delta\delta_{mn}\right] \Det^{-1/2}\left[1-2 i\left(-\Delta\right)^{-1}\ad (P^+F)_{mn} \right]
\end{split}
\end{equation}
The orbital contribution combines with the \emph{FP} determinant to give:
\begin{equation}
\Det^{-1/2} \left[-\Delta\delta_{mn}\right] \Delta_\textrm{FP} \sim \exp\left[-\frac{N}{3\left(4\pi\right)^2} \log\left(\frac{\Lambda}{M}\right)\,S_\text{YM}\right]
\end{equation}
The spin term in the Jacobian gives only the $SD$ contribution of the spin term to the beta function in the original variables Eq.~\eqref{eq:standard}, because of the projector $P^+$:
\begin{equation}
\begin{split}
\Det^{-1/2}\left[1- 2 i\left(-\Delta\right)^{-1}\ad (P^+F)_{mn} \right] \sim \exp\left[\frac{4N}{\left(4\pi\right)^2} \log\left(\frac{\Lambda}{M}\right)\,  \int d^4x\, \tr (P^+F)_{mn}^2  \right]
\end{split}
\end{equation}
The $ASD$ part of the spin contribution is furnished by the extra determinant due to the nonlinear coupling to the external source computed in Eq.~\eqref{eq:detsource}. Were this additional contribution be absent, the beta function would not coincide with the one in the original variables.
Finally, employing the identity $S_\text{YM}=  \int d^4x \, [\tr (P^+F)_{mn}^2 + \tr (P^-F)_{mn}^2]  $, the local part of the effective action is:
\begin{equation}
\exp\left({-\Gamma}\right) = \exp\left[-\frac{1}{2g^2}S_\text{YM}+\frac{11N}{3\left(4\pi\right)^2} \log\left(\frac{\Lambda}{M}\right)\,S_\text{YM}\right]
\end{equation}
that agrees perfectly with the standard result Eq.~\eqref{eq:standard}.
The \emph{YM} one-loop beta function follows:
\begin{equation}
\beta(g) = \frac{\partial g}{\partial\log \Lambda} = -\frac{11Ng^3}{3\left(4\pi\right)^2}
\end{equation}

\providecommand{\href}[2]{#2}\begingroup\raggedright\endgroup

\end{document}